\documentclass[preprint]{aastex}

\received{}
\accepted{}
\journalid{}{}
\articleid{}{}

\shortauthors{Carpenter, Heyer, \& Snell}
\shorttitle{Clusters in W3/W4/W5}

\newcommand{\ts}{\thinspace}
\newcommand{\simless}{\mathbin{\lower 3pt\hbox
     {$\rlap{\raise 5pt\hbox{$\char'074$}}\mathchar"7218$}}}
\newcommand{\simgreat}{\mathbin{\lower 3pt\hbox
     {$\rlap{\raise 5pt\hbox{$\char'076$}}\mathchar"7218$}}}

\newcommand{\about}    {$\approx$\ts}

\newcommand{\aboutmore}{$\simgreat$\ts}

\newcommand{\KP}{$K'$}

\newcommand{\M}{$^{\rm m}$}

\newcommand{\mic}{\ts\micron}

\newcommand{\msun}{\ts M$_\odot$}
\newcommand{\lsun}{\ts L$_\odot$}

\newcommand{\etal}{et~al.}

\newcommand{\co}{$^{12}$CO}
\newcommand{\coj}{$^{12}$CO(1--0)}
\newcommand{\thco}{$^{13}$CO}
\newcommand{\thcoj}{$^{13}$CO(1--0)}
\newcommand{\kkms}{\ts K\ts km\ts s$^{-1}$}
\newcommand{\kms}{\ts km\ts s$^{-1}$}
\newcommand{\kpc}{\ts kpc}
\newcommand{\pc}{\ts pc}
\newcommand{\tmb}{$T_{\rm MB}$}
\newcommand{\lhi}{$\lambda$\ts21\ts cm}
\newcommand{\hii}{H~{\sc ii}}

\def\insertplot#1#2#3#4#5#6#7{
\vskip 10pt\nobreak\hbox to \hsize{\hss\dimen0=#3in\hbox to #6\dimen0{%
\dimen0=#2in\vbox to #6\dimen0{\vss
\special{ps: plotfile #1}
\special{ps::[end]
  PGPLOT restore
}
}\hss}\hss}\vskip 10pt}

\begin{document}

\title{Embedded Stellar Clusters in the \\ W3/W4/W5 Molecular Cloud Complex}

\author{John M. Carpenter\\
        email: jmc@astro.caltech.edu}
\affil{California Institute of Technology, 
       Department of Astronomy, MS 105-24, \\ Pasadena, CA 91125}

\author{Mark H. Heyer\\
email: heyer@fermat.astro.umass.edu}

\and

\author{Ronald L. Snell\\
email: snell@fcrao1.astro.umass.edu}

\affil{University of Massachusetts, Department of Astronomy,
       Amherst, MA 01003}


\begin{abstract}

We analyze the embedded stellar content in the vicinity of the W3/W4/W5 \hii\ 
regions using the FCRAO Outer Galaxy \coj\ Survey, the IRAS Point Source 
Catalog, published radio continuum surveys, and new near-infrared and 
molecular line observations. Thirty-four IRAS Point Sources are identified 
that have far-infrared colors characteristic of embedded star forming regions,
and we have obtained \KP\ mosaics and \thcoj\ maps for 32 of them. Ten of the
IRAS sources are associated with an OB star and 19 with a stellar cluster,
although three OB stars are not identified with a cluster. Half of the 
embedded stellar population identified in the \KP\ images is found in just the 
5 richest clusters, and 61\% is contained in IRAS sources associated with an 
embedded OB star. Thus rich clusters around OB stars contribute substantially
to the stellar population currently forming in the W3/W4/W5 region.
Approximately 39\% of the cluster population is embedded in small clouds with 
an average mass of \about 130\msun\ that are located as far as 100\pc\ from 
the W3/W4/W5 cloud complex. We speculate that these small clouds are fragments 
of a cloud complex dispersed by previous episodes of massive star formation. 
Finally, we find that 4 of the 5 known embedded massive star forming sites in 
the W3 molecular cloud are found along the interface with the W4 \hii\ region 
despite the fact that most of the molecular mass is contained in the interior 
regions of the cloud. These observations are consistent with the classical 
notion that the W4 \hii\ region has triggered massive star formation along the 
eastern edge of the W3 molecular cloud.

\end{abstract}


\section{Introduction}

The equilibrium of molecular clouds is frequently perturbed by passages 
through spiral arms, cloud-cloud collisions, shocks from OB stellar winds and 
ionization fronts, molecular outflows, and other energetic forces. Depending 
on the local conditions, these events can either compress or disperse the 
molecular cloud, and consequently, induce or halt any future star formation
\citep{Elm92}. These interactions occur more frequently among 
clouds located in the disk of the Milky Way, and therefore potentially, the 
average star formation characteristics of clouds in the disk can differ from 
those that are far removed from the Galactic Plane. Since most of the 
molecular material is confined to the disk of the Galaxy \citep{CSS88,Dame87},
studying the stellar content of Galactic Plane molecular clouds is necessary 
to establish the conditions under which most star form.

The biggest challenge in studying molecular clouds in the disk of the 
Galaxy is that they are generally rather distant from the sun at several 
kiloparsecs and are often blended together in projection, especially in the 
inner Galaxy (see, e.g., Lee, Snell, \& Dickman 1990). Thus detecting and 
uniquely associating star formation sites within individual molecular clouds 
can often be problematic. To minimize these difficulties, the Five College 
Radio Astronomy Observatory (FCRAO) has recently completed a \co\ survey of 
the Outer Galaxy \citep{Hey98}. This survey encompasses 330\ts deg$^2$ 
of the second Galactic quadrant at subarcminute resolution and sampling, and 
represents the most detailed examination to date of the molecular interstellar 
medium. Most lines of sight in the survey contain a single molecular cloud, 
and moreover, this region of the Galaxy includes the closest approach of the 
Perseus Spiral Arm to the sun at a distance of \about 2\kpc\ \citep{GG76}. In 
these respects, this is one of the best regions in the Milky Way to 
investigate the stellar properties of Galactic Plane molecular clouds.

In this paper, we investigate the stellar characteristics of molecular clouds 
within the Perseus Spiral Arm, and more specifically, clouds in the vicinity 
of the W3/W4/W5 \hii\ regions, using the FCRAO Outer Galaxy \co\ Survey, the 
IRAS Point Source Catalog, published radio continuum observations, and new
near-infrared and molecular line data. The W3/W4/W5 chain of \hii\ 
regions is ionized by members of the Cas~OB6 association and extends over 
150\pc\ along the Perseus Arm. The winds and ionizing flux from the massive 
stars have clearly impacted the interstellar medium by creating a galactic
chimney out of the atomic hydrogen gas \citep{NTD96}, shaping molecular clouds 
into cometary globules with parsec sized tails \citep{Hey96}, and possibly 
inducing a second generation of OB star formation \citep{Lada78,Thronson80}. 
Despite the vigorous star formation activity in the past, a substantial mass 
of molecular gas remains \citep{HT98,Digel96, Lada78}. The W3 molecular cloud 
alone has \about 10$^5$\msun\ of molecular material spread over a \about 
60\pc\ region \citep{Deane00,Lada78} and is one of the most massive molecular 
clouds in the outer Galaxy \citep{Hey98}. The luminous star forming sites
W3 Main, W3(OH), W3 North, and AFGL~333 \citep{Thronson80}, a ridge of dense
molecular gas \citep{Tie98}, and the presence of embedded clusters 
throughout the W3/W4/W5 cloud complex \citep{Tie98,Deharveng97,Megeath96,
Hodapp94,Carp93} all attest to the continued star formation activity in this 
region. In addition to the massive W3 Giant Molecular Cloud (GMC), a number of 
small clouds with similar \coj\ velocities as the W3/W4/W5 region are found 
scattered throughout the area \citep{Hey98}. 

To investigate the 
most recent generation of star formation throughout the W3/W4/W5 region, we 
have selected a sample of likely embedded star forming sites using 
the IRAS point source catalog. In conjunction with new \thcoj\ and \KP\ 
band near-infrared observations, we investigate the spatial distribution of 
star forming regions, their associated molecular cloud properties, and the 
incidence of stellar clusters, and use these data to establish where most of 
the stars are now forming in the W3/W4/W5 region. These results are presented 
as follows. Section~\ref{obs} discusses the criteria used to identify star 
forming regions from the IRAS Point Source Catalog and describe the 
observations and data reduction procedures for the new molecular line and 
near-infrared surveys. In Section~\ref{results}, we characterize the 
stellar content associated with the IRAS sources based upon the far-infrared 
luminosity, published radio continuum observations, and the incidence of any 
stellar clusters detected in the \KP\ band mosaics. The implication of these 
results on star formation in the W3/W4/W5 region is discussed in 
Section~\ref{discussion}, and our conclusions are summarized in 
Section~\ref{summary}.

\section{Observations}
\label{obs}

The region analyzed for this study encompasses the area between galactic 
longitudes $\ell = 130\arcdeg$\ to $139\arcdeg$\ and latitudes $b = 
-2.2\arcdeg$\ to $+4.5\arcdeg$. Figures~\ref{w345_co} and \ref{w345_21cm}
show this region of the Galaxy as observed in \coj\ from the FCRAO Outer
Galaxy Survey \citep{Hey98} and in \lhi\ radio continuum emission from
the DRAO Galactic Plane Survey \citep{NTD97}. The W3, W4, and W5 \hii\ regions
and the associated molecular clouds are evident in these figures. A list of 
candidate embedded star forming regions within these clouds were selected 
using the IRAS point source catalog, and follow-up observations of these
sources were conducted to ascertain their stellar content. The selection 
criteria for the IRAS point sources and the ancillary observations are 
described below.

\subsection{IRAS Point Source Selection}
\label{iras}

Guided by the far-infrared colors of known embedded star forming regions
\citep{Carp91,Ken90,Bei86} and the properties of the IRAS sources observed by 
\citet{Wou90}, the following five criteria were used to identify embedded star 
forming regions in the IRAS Point Source Catalog, Version 2.1.
(1) The source has a high or moderate quality flux measurement at both 
       25\mic\ and 60\mic.
(2) The flux density at 25\mic\ is $\ge 0.4$\ts Jy.
(3) The flux density at 60\mic\ is $\ge 1.0$\ts Jy.
(4) The ratio of the 60\mic\ to the 25\mic\ flux density is $>$ 1.0.
(5) The ratio of the 100\mic\ to the 60\mic\ flux density is $\le$ 4.0.
The two flux density ratio criteria were designed to isolate sources with 
rising spectral energy distributions, but to eliminate the reddest objects 
that are often indicative of infrared cirrus. Any upper limits to the 
100\mic\ flux density were used as appropriate in evaluating the ratios. 

Of the sources meeting the above criteria, 34 were identified with a molecular
cloud that has a \co\ velocity between -57\kms\ $\le v_{\rm lsr} \le -32$\kms\ 
and are likely in the Perseus spiral arm \citep{Hey98}. These 34 sources
are listed in Table~\ref{tbl:iras}, along with the galactic and equatorial
coordinates, the velocity of the \coj\ emission coincident with the IRAS 
source, and any source identifications. All but two of these sources, as noted 
in Table~\ref{tbl:iras}, were imaged at \KP\ band and mapped in \thco\ as 
described below. The spatial distribution of the IRAS point sources is shown 
by the open circles in Figure~\ref{w345_co}, where the circle diameter is 
proportional to the far-infrared luminosity (see Section~\ref{irlum}). The 
source list does not include W3 Main, which is confused at 60\mic\ in the IRAS 
survey and did not meet the selection criteria. Near-infrared observations of 
this source have been presented by Megeath \etal~(1996; see also Tieftrunk
\etal~1998 and Hodapp~1994). Also, two of the IRAS sources have previously 
been associated with galaxies \citep{Wei80}. However, the strong \co\ emission 
and the apparent stellar cluster (see Section~\ref{clusters}) associated with 
these two sources suggests that the far-infrared emission originates from 
embedded stars and not an extragalactic object. 

\subsection{\KP\ Band Imaging}

Near-infrared mosaics in a \KP\ band filter \citep{WC92} were obtained for 32
of the 34 IRAS sources in Table~\ref{tbl:iras}. Time constraints prevented us
from observing IRAS 02081+6225 and 02204+6128. The images were obtained over 
a two night period in 1996 October using QUIRC at the University of Hawaii 
2.24\ts m telescope on Mauna Kea through thin cirrus with seeing conditions of 
\about 0.5-0.6\arcsec. QUIRC contains a 1024 $\times$ 1024 HgCdTe array and 
was used at the $f$/10 focus to provide a plate scale of 0.186\arcsec\ 
pixel$^{-1}$ and an instantaneous field of view of $3.2' \times 3.2'$. For 
each IRAS source, a 5\arcmin\ $\times$ 5\arcmin\ mosaic aligned in the 
equatorial coordinate system was obtained that consists of 12 dithered frames 
with an exposure time of 30 seconds per frame. Sky frames were constructed by
median filtering images free of extended nebulosity.
The sky-subtracted frames were then corrected by a flat field 
image derived from a series of exposures of the dome interior with and without 
illumination from incandescent lights. In constructing the mosaics, three 
frames were coadded per pixel position in the mosaic in order to maintain a 
constant noise level across the final image. Coadded pixels near the edge of 
the mosaic that have only one or two observations were discarded. Astrometry
for 23 of the 32 mosaics (see Table~\ref{tbl:iras}) were established using 
images from the 2 Micron All Sky Survey (2MASS). 2MASS images for the 9 
remaining mosaics were not available at the time of this study, and we assumed 
that the center of these mosaics corresponds to the IRAS point source position.
Based on the results from registering the 23 mosaics with available 2MASS 
images, we expect that the astrometry for these 9 mosaics to be accurate to be 
$\le$ 30\arcsec.

Stars were identified in the mosaics using DAOFIND in IRAF. The noise for each 
mosaic was measured empirically, and a 5$\sigma$ detection threshold was used 
to create an initial source list. All mosaics were then visually inspected 
to remove a few saturated stars and any obvious non-stellar objects (e.g. 
ghosts, nebulosity knots), and to add any stars that were not identified by 
DAOFIND. Photometry was performed using the point-spread fitting task DAOPHOT 
in IRAF. The point spread function (PSF) was determined for each mosaic using 
several bright, isolated stars in the image. After fitting the PSF to each 
star in the point source list and subtracting the fit from the mosaic, the 
resulting image was examined for any additional point sources that were 
initially missed due to source confusion. These stars were added to the 
detection list and were also measured using DAOPHOT. Finally, objects 
identified by DAOPHOT as being unusually extended based upon the ``sharp'' 
statistic (e.g. galaxies, nebulosity) were removed form the point source list. 
The number of extended objects removed by this criteria amounted to less than 
10\% of the total number of sources identified. The photometry was calibrated 
by observing standard stars in the UKIRT faint standards list \citep{Cas92} 
and assuming a \KP\ band extinction coefficient of 0.08 mag/airmass. The RMS 
scatter over the two nights in the photometric zero points derived from the 
standard observations is 3\%.

The differential completeness limit of the survey was established by adding 
artificial stars of known magnitude to one of the mosaics and determining 
the fraction of the stars that could be recovered at the 5$\sigma$ detection 
threshold. When adding artificial stars, care was taken not to add objects 
near a known star or within nebulous regions. The completeness limit in these
more confused regions will obviously occur at a brighter magnitude. 
Approximately 90\% of the stars with a \KP\ magnitude of 17.5\M\ were
recoverable in this automated procedure. The star counts analyzed in the paper 
are therefore for objects with \KP\ magnitudes between 11.5\M\ (the saturation
limit) and 17.5\M.

\subsection{\thco\ Mapping}

A region of size ($\Delta\ell \times \Delta b$) = ($6.2' \times 4.8'$) toward 
the 32 IRAS point sources was mapped in \thcoj\ (110.201370\ts GHz) using the 
SEQUOIA receiver array on the 14\ts m telescope operated by the Five College 
Radio Astronomy Observatory (FCRAO) in the spring of 1998.
At the time of the observations, SEQUOIA had 12 pixel elements. 
The full-width-at-half-maximum (FWHM) beam size of the FCRAO telescope at the 
observed frequency is 47$''$, and the maps were sampled every 22$''$, or 
approximately the Nyquist sampling interval. The backends for each pixel in 
the SEQUOIA array consisted of an autocorrelator spectrometer configured to 
achieve a velocity resolution of 0.064\kms\ over a 54\kms\ velocity interval. 
The data were obtained in frequency switching mode with an offset of 4~MHz 
between the nominal and reference frequency. The \thco\ data presented here 
have been corrected by the main beam efficiency, previously measured to be 
$\eta_{\rm B}$ = 0.45. The RMS noise is typically $\Delta$\tmb\ = 0.7-1.1~K 
per channel.

\section{Results}
\label{results}

\subsection{Images}

Molecular line and \KP\ band images of the 32 IRAS point sources observed
for this study are shown in Figure~\ref{images}. Four images are shown for
each IRAS source: 
  (1) a \coj\ integrated intensity map from the FCRAO Outer Galaxy Survey
      over a $30' \times 30'$ area centered on the IRAS point source position 
      (far left panels),
  (2) the \thco\ integrated intensity map over a $6.2' \times 4.8'$ region,
  (3) the \KP\ band mosaic, and
  (4) the \KP\ stellar surface density map.
The \co\ images are shown over a larger extent than the \thco\ maps and \KP\ 
band mosaics in order to place the IRAS source in context of the large scale 
molecular cloud in the region. Figure~\ref{images} indicates that most of the 
\thco\ maps peak near the IRAS point source positions and suggests that the 
IRAS sources are indeed related with the molecular gas.
In the remainder of this section, we characterize the stellar population 
associated with these IRAS sources as inferred from the far-infrared 
luminosity, the presence of an embedded massive star as indicated by radio 
continuum emission, and the identification of stellar clusters from \KP\ 
band star counts.

\subsection{Far-Infrared Luminosities}
\label{irlum}

The stellar content associated with the IRAS sources can be constrained to 
first order by assuming that dust absorbs a substantial fraction of the 
stellar bolometric luminosity and re-emits the radiation in the far-infrared.
Since many of the IRAS sources are associated with a cluster of stars (see 
Section~\ref{id_clusters}), the far-infrared emission actually sets a limit
on the most massive star that may be forming in these regions. The luminosity 
emitted in the 12\mic, 25\mic, 60\mic\ bands (L$_{FIR}$) was computed by 
summing the observed flux densities in the individual IRAS band passes using 
the formula
\begin{eqnarray}
 \rm{L_{FIR}} & = & \rm{4\pi D^2 \sum_i(S_{\nu i}\;\Delta\nu_i)}\nonumber\\
              & = & \rm{0.30~\Bigl({D\over kpc}\Bigr)^2~\sum_i
                    \Bigl({S_{\nu i} \over Jy}\Bigr)\:
                    \Bigl({\Delta\nu_i\over 10^{12}\;Hz}\Bigr)~L_\odot},
\end{eqnarray}
where D is the distance to the source in kiloparsecs (assumed to be 2.35\kpc;
Massey, Johnson, \& DeGioia-Eastwood 1995), $\Delta\nu$ the IRAS band width, 
and S$_\nu$ the observed flux density. The sum does not extend over the
100\mic\ band since many sources are confused at 100\mic\ and this band
could not be applied consistently for the entire sample. Thus the computed 
far-infrared luminosities (see column 2 in Table~\ref{tbl:clusters}) will 
underestimate the actual bolometric and far-infrared luminosities. For IRAS 
sources in the W3/W4/W5 region that do have high quality 100\mic\ detections, 
we found that the 12\mic-60\mic\ luminosity underestimates the total IRAS 
far-infrared luminosity by \about 30\% on average.

\begin{figure}
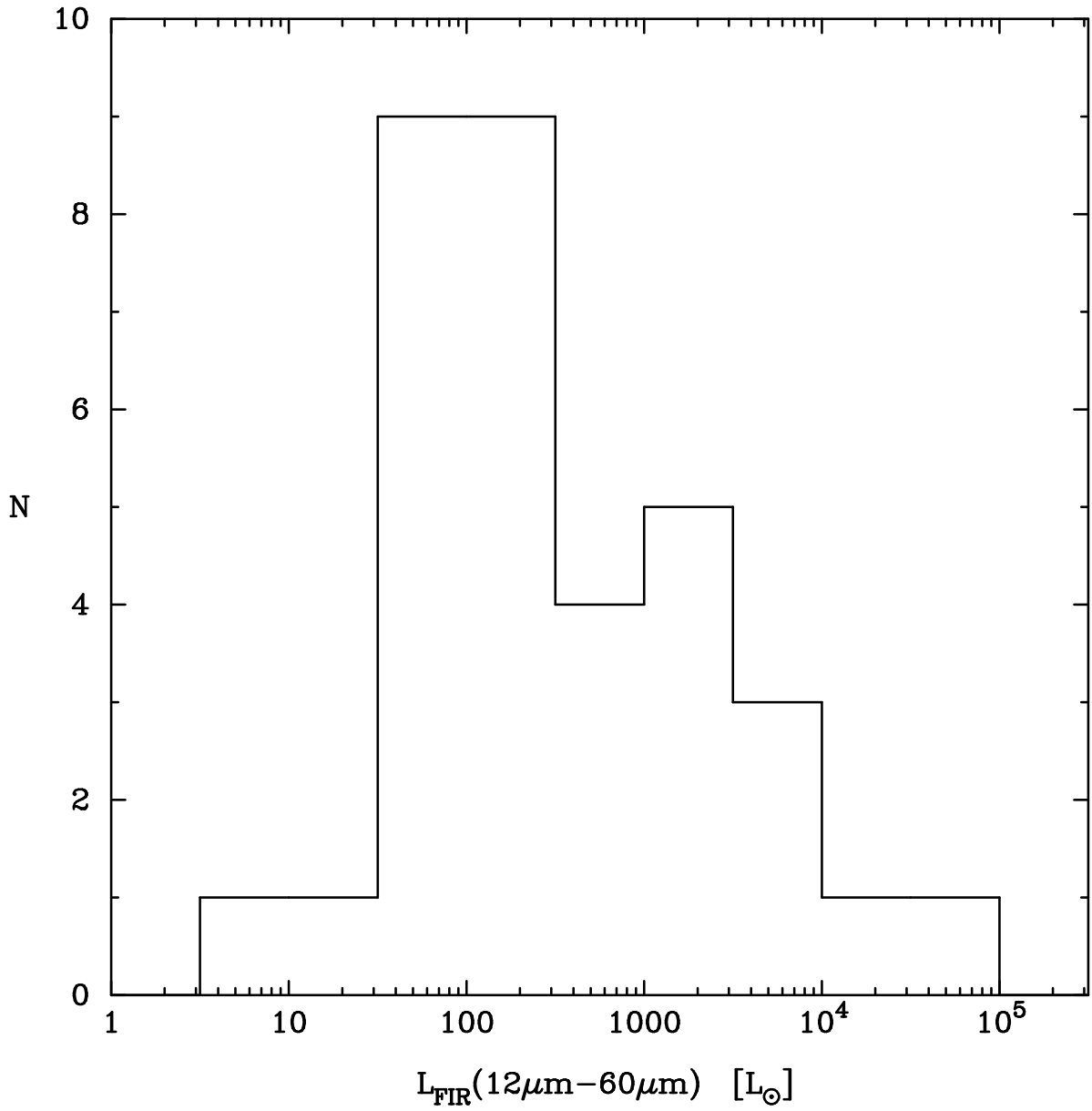

\insertplot{figure4.ps}{6.0}{4}{0.0}{3.2}{0.9}{1}
\caption
{
    Histogram of the far-infrared luminosity in the 12\mic, 25\mic, and 60\mic\
    IRAS bands for the 34 IRAS sources in our sample.
    \label{lfir}
}
\end{figure}

A histogram of the derived far-infrared luminosities for the 34 
IRAS sources is shown in Figure~\ref{lfir}. The observed luminosities range
between 9\lsun\ and 46,000\lsun\ with the peak of the distribution at \about 
100\lsun. The brightest sources have luminosities similar to that of early B 
type zero age main sequence (ZAMS) stars \citep{Pan73} assuming that most of 
the far-infrared luminosity originates from a single object. The lowest
luminosities in the histogram are a result of the selection criteria. The 
flux density criteria alone used to select the IRAS sources implies a detection 
limit of 8\lsun. Further, since the 60\mic\ flux density is on average 11 
times larger than the 25\mic\ flux density for sources in our sample, the 
detection limit set by the spectral energy distribution and flux density 
limits is 23\lsun. The decline in the number of sources with luminosities 
fainter than \about 100\lsun\ then is likely a result of incompleteness in the 
IRAS point source catalog. This 100\lsun\ limit corresponds to a 1 Myr, 
3\msun\ pre-main-sequence object \citep{PS93}, or a late B ZAMS star.

\subsection{Radio Continuum Emission}
\label{continuum}

The stellar content associated with the IRAS sources can be further constrained
by using radio continuum observations to estimate the spectral type of the 
most massive star.  The list of radio continuum sources in the W3/W4/W5 region 
were taken primarily from the $\lambda$\ts20\ts cm NRAO/VLA Sky Survey (NVSS; 
Condon \etal~1998), but also published targeted observations 
\citep{KCW94,Mcc91,Carp91}. Sources in the NVSS catalog were deemed associated 
with an IRAS source if the radio and far-infrared coordinates agreed to within 
30\arcsec. While no spectral information is available from the NVSS catalog to 
confirm that these objects are actually compact \hii\ regions, the random 
probability of a false association between an extragalactic radio continuum 
source and the IRAS point source is only \about 0.01 for the adopted 30\arcsec\ 
matching radius \citep{Con98}. The observed radio continuum flux was used to 
estimate the number of ionizing photons (see, e.g., Carpenter, Snell, \& 
Schloerb 1991) and consequently the spectral type of the ionizing star 
\citep{Pan73}. The inferred spectral types and references for the radio 
continuum observations are provided in columns 3 and 4 of 
Table~\ref{tbl:clusters}. Ten of the 32 IRAS point sources have radio 
continuum detections, and the inferred spectral types range from B2 ZAMS to O7 
ZAMS. If the stellar parameters from \citet{VGS96} are used instead of 
\cite{Pan73} to infer the spectral type, all but one of the detected sources 
will have a spectral type later than B0.5 ZAMS.

Two of the IRAS sources (IRAS 01546+6319 and 02511+6023) associated with a B 
type star as based on their radio continuum flux have a 12\mic-60\mic\ 
luminosity that is 2 orders of magnitude less than that expected for such a 
massive star \citep{Pan73}. Either the association between these radio 
continuum and IRAS sources is incorrect and the radio continuum source is an 
extragalactic object, or a substantial fraction of the bolometric luminosity 
is radiated at shorter wavelengths. The latter situation may occur if a star 
forming region is relatively evolved and the circumstellar dust no longer 
completely absorbs and re-emit the stellar radiation, although no obvious 
bright star is present in the \KP\ mosaics or on the Palomar Observatory Sky 
Survey prints to indicate that this may be the case. Nonetheless, we assume in 
the remainder of this paper these two IRAS sources are indeed massive star 
forming regions. Our general conclusions will not change if this assumption is 
incorrect.

\subsection{Identification of Stellar Clusters}
\label{id_clusters}

As another means to characterize the stellar content associated with the IRAS 
Point Sources, we searched for stellar clusters in the \KP\ images using the 
procedure adopted by Carpenter, Snell, \& Schloerb (1995; see also Carpenter
\etal~1997). Briefly, histograms of the stellar field star density were 
generated for each mosaic using 20$''\times$20$''$ counting bins sampled every 
10$''$. The observed frequency distribution of counts at low stellar surface 
densities in these histograms usually resembles a Poisson distribution, which 
is identified with field stars and embedded stars randomly distributed across 
the \KP\ band mosaic. By fitting a Poisson distribution to these lower surface 
density bins, the mean stellar surface density of randomly distributed stars 
can be determined. Stellar surface density bins that significantly exceed this 
mean surface density are identified as possible clusters. Contour maps of the 
stellar surface density are shown in Figure~\ref{images} for each of the 
mosaics. The lowest contour in each map begin at 2$\sigma$ for a Poisson 
distribution above the mean stellar surface density with 3$\sigma$ contour 
intervals. 

A cluster was identified in Figure~\ref{images} if the total number of stars 
within the 2$\sigma$ contour represents a 5$\sigma$ enhancement with respect 
to the expected stellar background level. Further, we require that the 
identified cluster be near the center of the mosaic and the IRAS point source 
position. In a few instances, apparent clusters were identified near the edge 
of the mosaic. In nearly all of these cases, the extinction through the cloud 
was large enough and variable that a Poisson distribution is a poor 
representation of the star counts, and these clusters most likely represent 
regions within the image where the extinction becomes low. Any such 
``cluster'' is almost certainly a projection of unrelated field stars and was 
excluded from the final cluster list.

Table~\ref{tbl:clusters} summarizes the properties of the identified clusters,
including the effective cluster radius ($R_{\rm eff}$), the number of stars 
observed within the 2$\sigma$ boundary (N$_{\rm s}$), and the number of 
stars inferred for the cluster after subtracting off the expected field star 
population (N$_{\rm cluster}$). The effective radius is defined as 
$\sqrt{A/\pi}$, where $A$ is the area within the 2$\sigma$ contour. Of the 32 
IRAS sources, 19 have identifiable clusters, with the number of cluster 
members ranging from 20 to 240 stars. Our results agree with previous studies
in that clusters are identified around IRAS~02232+6138 \citep{Tie98},
IRAS~02575+6017 \citep{Deharveng97,Hodapp94,Carp93}, and IRAS~02593+6016 
\citep{Carp93}. However, \citet{Tie98} visually noted a second cluster near 
IRAS~02232+6138 that is within the field of view of our mosaic. This grouping
of stars does not meet the surface density criteria adopted here to be 
identified as a cluster.

The cluster membership listed in Table~\ref{tbl:clusters} are lower limits to 
the actual stellar population in these regions, mainly due to the finite 
sensitivity of the observations. To compare these clusters with other star 
forming regions, we computed the fraction of stars in nearby embedded clusters 
that would be detectable at the distance of W3 considering both differences in 
sensitivity and resolution. In the UKIRT survey of the MonR2 cluster by 
\citet{Carp97}, 246 of 378 stars (65\%) would be detectable with our QUIRC
observations at the distance of W3/W4/W5. Similarly, we could detect
\about 410 (53\%) of the \about 780 stars in the inner $5'\times5'$ of the 
Orion Nebula Cluster \citep{Hillenbrand00}, and \about 75\% of the 94 stars 
brighter than $m_k$ = 14.5\M\ in the NGC\ts1333 cluster \citep{Lada96}. 
Assuming that the extinction toward the W3/W4/W5 clusters is not substantially 
different than toward these comparison regions, the richer W3/W4/W5 clusters 
(\about 200 stars) are comparable to MonR2, but not as rich as the Orion 
Nebula Cluster. Several of the W3/W4/W5 clusters though have significantly 
fewer stars than these comparison regions.

\subsection{Molecular Cloud Properties}

The properties of the molecular clouds associated with the IRAS sources as
derived from \thcoj\ emission are summarized in Table~\ref{tbl:clouds}. The 
cloud sizes are defined as the circular radius needed to produce the area 
within the FWHM integrated intensity contour. The cloud masses were computed 
from the integrated \thcoj\ intensity by assuming that the emission is 
optically thin and that the \thco/H$_2$ abundance is $1.5 \times 10^{-6}$
\citep{BC86}. Statistical equilibrium calculations indicate that for H$_2$ 
volume densities between 300\ts cm$^{-3}$ and 10$^4$\ts cm$^{-3}$ and kinetic 
temperatures between 10\ts K and 20\ts K, the fraction of the \thco\ molecules 
in the J=1 rotational state varies between 0.41 to 0.54, and 0.48 was adopted 
as a typical value. A factor of 1.36 was included in the calculations to 
include the mass contribution from helium and other elements. The \thco\ 
integrated intensity used to calculate the masses includes the region within 
the FWHM intensity contour, but was also extrapolated to include emission 
outside this contour level by assuming a gaussian intensity distribution. 
Lower limits to the cloud sizes and masses are reported for sources in which 
the \thco\ maps did not completely encompass the half power contour level. The 
derived cloud masses range from 27\msun\ to $>$ 3200\msun\ (see 
Table~\ref{tbl:clouds}). Many of these clouds, and in particular the 
more massive objects, are parts of the extended W3 and W5 GMCs. Several of the 
clouds are small and distant from any GMC (see Figure~\ref{w345_co}) and are 
nearly completely mapped with these \thco\ observations.

\section{Discussion}
\label{discussion}

\subsection{Stellar Clusters in W3/W4/W5}
\label{clusters}

\begin{figure}
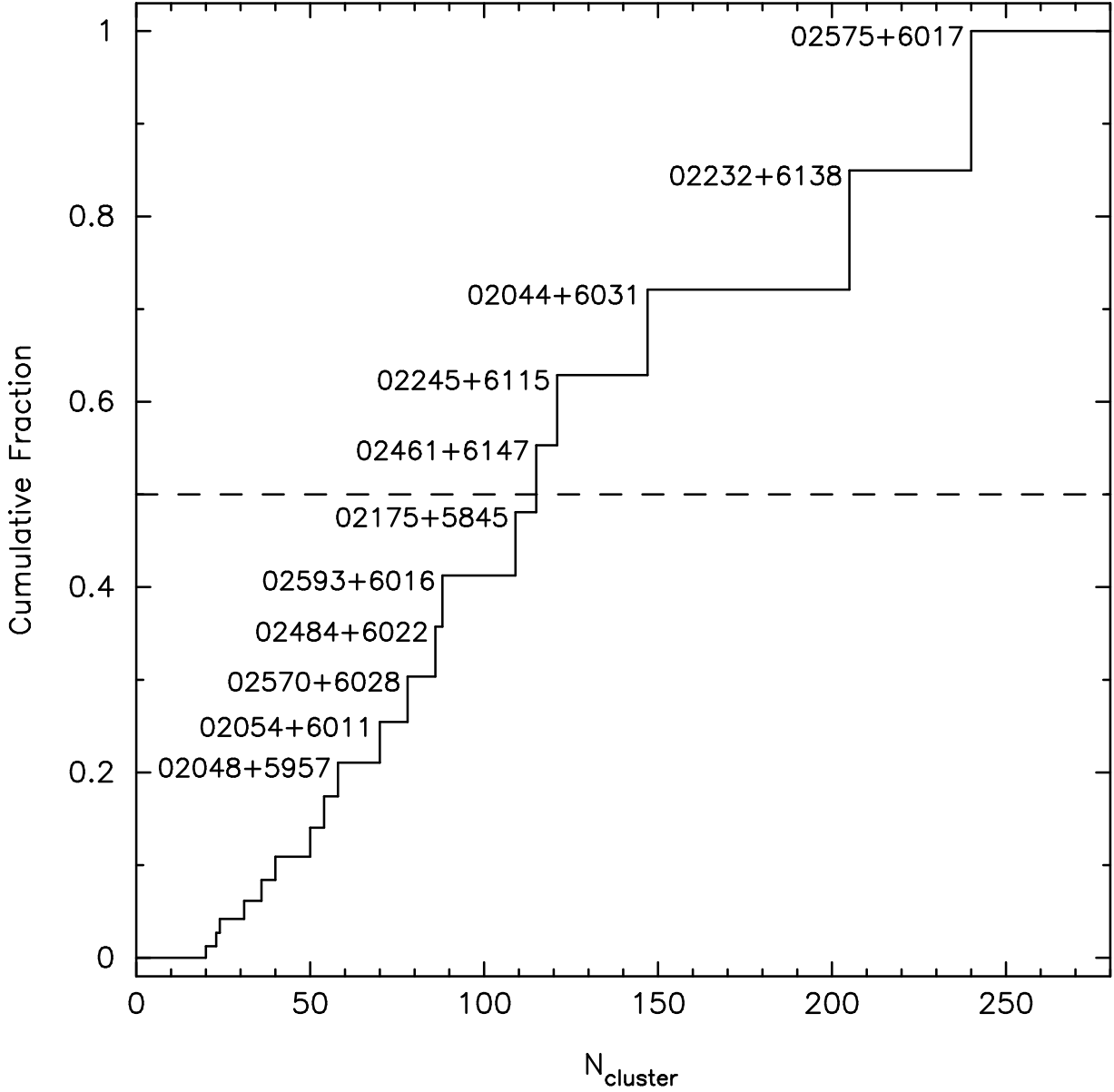

\insertplot{figure5.ps}{6.0}{4}{0}{3.2}{0.9}{1}
\caption
{
    The cumulative distribution of the total number of stars identified in
    clusters as a function of the number of stars in an individual cluster.
    The star counts have been normalized by the total cluster population
    (1595 stars). The richest clusters are labeled by their IRAS point source
    name, and the horizontal dashed line shows the 50\% point in the cluster
    population. These results indicate that half of the detected cluster
    population is found in just the five richest clusters.
    \label{cum}
}
\end{figure}

From the results presented in Section~\ref{results} we can begin to assess 
the properties of the regions that are forming most of the cluster stellar
population. The 19 identified clusters contain a total of 1595 stars within 
the 2\ts$\sigma$ boundary after subtracting off the expected field star 
contamination. (If the 1\ts$\sigma$ boundary is used to define the clusters, 
the total membership increases by 33\%.) Figure~\ref{cum} shows the normalized 
cumulative distribution of the total number of stars in all 19 clusters as a 
function of the number of stars in an individual cluster. As Figure~\ref{cum} 
shows, 52\% of the cluster members are found in just the 5 richest clusters. 
Even if we assume that the 13 IRAS sources without identified clusters have 20 
stars each (the smallest population cluster identified in this survey), the 5 
richest clusters would still contain 45\% of the total cluster membership. 
These results do not change significantly if we include the W3 Main cluster, 
which has \about 87 stars to a comparable sensitivity limit as our survey 
\citep{Megeath96}, or the other clusters identified along the W3 ridge
\citep{Tie98}. 

We can also investigate the fraction of the cluster population associated 
with embedded OB stars. The 5 rich clusters described above that contain the 
majority of the cluster population are each associated with a massive star, 
and in total, clusters around OB stars contain 61\% of the cluster population.
Not all of the embedded OB stars though are associated with clusters. IRAS 
02230+6202, 02511+6023, and 02531+6032 do not show a significant enhancement 
in the stellar surface density despite having a radio continuum detection. 
Based on the field star initial mass function, one would expect an OB type
star to form along with a few hundred lower mass objects \citep{MS79}. It 
would be remarkable then if these regions are truly forming a single massive 
star, and would imply that the stellar mass function in these regions is 
strongly skewed toward high mass objects. With high spatial resolution data at 
only one wavelength, however, we cannot yet rule out the possibility that the 
lack of a cluster in these regions is merely due to a large amount of 
extinction, and additionally in the case IRAS~02230+6202, a bright nebular 
background (see Fig.~\ref{images}), that prevents us from detecting the 
underlying cluster. Deeper observations in the near- and mid-infrared should 
be able to establish more definitively if these OB stars are truly forming in 
isolation.

In principle the above results could change if a large number of low 
luminosity IRAS sources exist that did not meet our selection criteria but 
contribute substantially to the cluster population. Based on the peak of the
histogram in Figure~\ref{lfir}, we assume that the luminosity completeness 
limit for the IRAS point sources is \about 200\lsun, and a power law function 
was fitted to the histogram bins more luminous than this limit. Extrapolation 
of this power fit suggests that there could be \about 40 additional sources 
between 10\lsun\ and 200\lsun\ that are not already in our sample.
Even if we assume these sources have 20 stars each (the smallest cluster size 
detected here), the IRAS sources associated with OB will still contain more 
than a third of the total cluster population. These suggest indicate that the 
formation of dense stellar clusters surrounding OB stars represent a 
significant component of the cluster population in the W3/W4/W5 region. 

Our results for the W3/W4/W5 region can be compared with other near-infrared
surveys of molecular clouds. The regions that have been most extensively 
studied on a global scale are the Orion~A and Orion~B molecular clouds, and
in both of these clouds, dense stellar clusters contribute significantly to 
the total young stellar population. In Orion~B, approximately 96\% of the 
stellar population is found in just 4 clusters \citep{Lada91,Li97}. Similarly, 
in Orion~A, at least 60\% of the total stellar population is found within the 
Orion Nebula Cluster alone \citep{Allen00,Meyer00}, while the rest is 
distributed more uniformly throughout the molecular cloud and in several 
small clusters \citep{Strom93}. Near-infrared surveys of other molecular 
clouds are not as extensive as those in Orion and are biased toward known star 
forming regions, but nevertheless, observations of NGC\ts2264 
\citep{Lada93,Piche93} and NGC\ts1333 \citep{Lada96} also suggest that at 
least half of the stars within these regions form within clusters as opposed 
a more uniformly distributed population (see also the review by Clarke, 
Bonnell, \& Hillenbrand~2000). While our survey is also not sensitive to any 
pervasive distributed stellar population, our results for the W3/W4/W5 region 
are similar to Orion in that a significant fraction of the cluster population 
is confined to just a few rich clusters. 

\subsection{Global Star Formation Characteristics}
\label{global}

Of the 32 IRAS sources in our sample that were mosaicked at \KP\ band, 21 are 
located in projection against the W3/W4/W5 \hii\ region/molecular cloud complex 
(see Figure~\ref{w345_21cm}). The remaining 11 sources, or 34\% of the sample, 
are separated from the W3/W4/W5 complex by as much as 100\pc. The molecular 
clouds associated with these 11 sources have radii of \about 0.5\pc\ with 
masses ranging from 27\msun\ to 410\msun, with an average mass of 130\msun. 
This contrasts with the massive star forming sites found in W3 that have cloud 
masses upwards of a few thousand solar masses or more. 

Despite their relatively small mass, the 11 isolated clouds are not devoid of 
star formation nor is their stellar population limited to low mass stars.
Eight of the 11 sources are forming a stellar cluster (see 
Table~\ref{tbl:clusters}), and 3 sources (IRAS 01546+6319, 02044+6031, and 
02561+6147) are associated with an early B type as indicated by the radio 
continuum emission. The fraction of the total cluster population found within 
these 11 clouds is 629/1595 = 39\% despite the fact they have only \about 11\% 
of the total molecular mass as derived from the \thco\ observations. These 
percentages are likely upper limits since it is more difficult to identify 
IRAS point sources (and consequently clusters) near W3 than in more isolated
regions.  Indeed, the W3 Main cluster, two clusters a few arcminutes from 
W3(OH), and a cluster around BD~+61~411 \citep{Tie98,Megeath96} are 
found along the W3 ridge but were not identified with the IRAS point source 
selection criteria adopted in this paper. From visual inspection of the 
$K$ band image presented in \citet{Tie98}, we estimate that these 
clusters in total may contain a few hundred stars. Even after accounting
for these stars, the isolated clouds contribute a non-negligible fraction to 
the total cluster population currently forming in the W3/W4/W5 region.

The presence of massive stars and stellar clusters in these low mass clouds 
is somewhat surprising  in that such characteristics are generally 
associated with massive GMCs. An interesting question is then how these 
isolated, star forming clouds originated. While the observations presented 
here cannot answer this question directly, we can speculate on their origin 
based upon the distribution of clouds in the vicinity of the W3/W4/W5 \hii\
regions. The correspondence between the \co\ and radio continuum emission 
shows that the \hii\ regions have re-shaped the spatial distribution of 
molecular gas in some instances (see Figure~\ref{w345_21cm}). For example,
the molecular gas near ($\ell,b$) \about (135\arcdeg,0\arcdeg) wraps around 
the southern edge of the W4 \hii\ region, and the cometary globule near 
($\ell,b$) \about (134.8\arcdeg,1.3\arcdeg) possibly formed from the 
interaction between the molecular gas and radiation pressure from OB stars in 
the W4 \hii\ region \citep{Hey96}. The apparent effect of these interactions 
is that the small, dense globules of gas (which were either pre-existing or 
were formed by radiation pressure) remain for longer time scales than the more 
diffuse molecular material which is ionized or dispersed. Such globules are 
prominent in other OB associations and evolved \hii\ regions as well, such as 
Orion~OB~1 \citep{Ogura98} and IC\ts1396 \citep{Patel98}. It is tempting to 
speculate then that the isolated clouds tens of parsecs away from W3/W4/W5 are 
the last remnants of a GMC that has been dispersed by OB stars. In fact, 
the Per OB1 association occupies the area south of the W3/W4/W5 between 
$\ell$ = 130\arcdeg\ to 138\arcdeg and $b = -$5\arcdeg\ to $-1$\arcdeg\ where 
several of the low mass clouds are found and is at nearly the same distance as 
the W3/W4/W5 \hii\ regions \citep{Garmany92}.

\subsection{Massive Star Formation in W3}
\label{w3}

The W3 region has long been proposed as a classic example of induced or
``triggered'' star formation in molecular clouds. In this scenario, shocks 
arising from ionization fronts compress the ambient molecular material such 
that a gravitational instability develops in the post-shocked gas, leading to 
gravitational collapse and the formation of a new generation of stars. In the 
specific case of the W3 molecular cloud, it has been suggested that the
expansion of the W4 \hii\ region triggered the formation of W3~Main, W3(OH),
W3~N, and AFGL~333, \citep{Lada78,Thronson80}. Alternatively, it has been 
proposed that diffuse \hii\ regions appear adjacent to embedded star forming 
sites because the associated dense molecular material has effectively slowed 
the expansion of the \hii\ regions, and thus no cause and effect relationship 
exists between the \hii\ regions and the newly formed stars. These 
two possibilities can be investigated with the data obtained for this study. 
If triggering is an important manner in which massive stars form, then 
embedded star forming regions throughout the W3/W4/W5 region should 
preferentially be located along the interfaces between ionization fronts and 
molecular clouds. Alternatively, if the expanding \hii\ regions have had 
minimal impact on the massive star formation activity, then massive star 
forming regions should be located randomly throughout the W3 molecular cloud. 
\citet{Thronson80} cited similar arguments in suggesting that massive star 
formation has indeed been triggered in the W3 region, and we re-investigate 
this issue here using the more extensive and sensitive observations provided 
by IRAS and recent radio continuum surveys.

Of the molecular clouds in the W3/W4/W5 region, only the W3 molecular cloud
extends tens of parsecs beyond the \hii\ regions to provide a clear 
distinction between IRAS sources located near and distant from the ionization 
fronts as seen in Figures~\ref{w345_co} and \ref{w345_21cm}. (Most of the IRAS
sources in the W5 molecular cloud are located along the W5 \hii\ region,
consistent with the triggered star formation hypothesis. However, unlike the
W3 region, the small extent of the W5 molecular cloud does not provide a 
``control'' field unaffected by the ionization front.) The W3 molecular cloud 
extends for \about 60\pc\ and contains \aboutmore 10$^5$\msun\ of molecular 
material \citep{Deane00,Lada78}. Of the IRAS sources in our sample, the three 
most luminous sources are found along the interface between the W4 \hii\ 
region and the W3 molecular cloud. This interface also includes the W3 Main 
(IRAS 02219+6152) massive star forming region \citep{Megeath96}, which is not 
in our sample since this object is confused at 60\mic\ in the IRAS survey. By 
contrast, in the interior of the W3 cloud (i.e. west of the W4 ionization 
front), no IRAS sources are found that meet our selection criteria and have a 
far-infrared luminosity in excess of 500\lsun.

Since W3 Main was not picked out by our IRAS selection criteria, we examine 
the possibility that massive star forming regions elsewhere in the W3 
molecular cloud may also have been missed. Using the IRAS point source
catalog, all sources were examined that have far-infrared luminosities
in excess of 200\lsun\ but have low quality flux densities at 25\mic\ and/or 
60\mic.  Where appropriate, the reported upper limits were used in estimating 
the luminosity. All sources more luminous than 200\lsun\ were considered since 
two of the IRAS point sources in our sample are associated with a B type star 
despite having a low far-infrared luminosity (see Section~\ref{continuum}). 
Using this search criteria, 12 additional sources were picked out in the W3 
region. Eight of these sources are near W3 Main, W3(OH), or AFGL 333 and are 
confused at 60\mic\ and 100\mic, and one source is found midway between AFGL 
333 and W3(OH). The other 4 IRAS sources are found interior to the cloud and 
are at least 30\arcmin\ (20 parsecs) from the W4 ionization front. Each of 
these four sources have far-infrared luminosities less than 430\lsun, and 
none are associated with a NVSS radio continuum point source \citep{Con98}. 

Recently, \citet{BKM00} identified a massive O8 star that is ionizing an 
extended \hii\ region near the southwestern edge of the W3 molecular cloud at 
($\ell,b)$ \about (133.425\arcdeg,0.055\arcdeg).
The source is associated with extended far-infrared emission and is not listed
as an IRAS 60\micron\ point source, and possibly may represent a more evolved
star forming region than the point sources analyzed here. This source is 
distant from the W4 ionization front and may indeed be an example of a massive 
star formation region that has formed spontaneously \citep{BKM00}. Nonetheless,
for the W3 molecular cloud as a whole, we conclude that the massive star 
forming regions are located preferentially (but not exclusively) along the W4 
ionization front and are not randomly distributed throughout the cloud. This 
is quite remarkable considering that \about 75\% of the molecular mass in the 
W3 molecular cloud (based on the observed \coj\ integrated intensity) is 
contained within the extended molecular gas component. These observations are 
thus consistent with the conjecture that massive star formation along the 
eastern edge of the W3 molecular cloud has been triggered by the W4 ionization 
front as previously proposed by \citet{Lada78} and \citet{Thronson80}.

\section{Summary}
\label{summary}

We have analyzed the star formation properties of molecular clouds in the
vicinity of the W3/W4/W5 \hii\ regions using the FCRAO Outer Galaxy \coj\ 
Survey, the IRAS point source catalog, published radio continuum observations, 
\KP\ band imaging, and \thcoj\ observations. We use these data to identify 34 
IRAS point sources in the Perseus spiral arm that have far-infrared colors 
characteristic of embedded star forming regions. The stellar population 
associated with 32 of these IRAS sources are investigated based upon 
far-infrared and radio continuum emission and the identification of stellar 
clusters in the \KP\ band mosaics. Our main conclusions are:

\begin{enumerate}
\item Nineteen stellar clusters are identified that contain a total of 1595 
      stars with \KP\ magnitudes between 11.5$^{\rm m}$ and 17.5$^{\rm m}$. 
      Approximately half of the total cluster population is found within the 5 
      richest clusters. Further, IRAS sources associated with embedded OB 
      stars as traced by radio continuum emission contain 61\% of the cluster 
      population. Thus clusters around OB stars contribute substantially to 
      the stellar population currently forming in the W3/W4/W5 region.

\item Eleven of the 32 IRAS point sources are located as far as 100\pc\
      from the W3/W4/W5 \hii\ region/molecular cloud complex. Despite the low 
      average cloud mass (130\msun) associated with these 11 sources, three of 
      these objects are associated with an OB star and 8 are forming a stellar 
      cluster. These 11 clouds contain \about 39\% of the total stellar 
      population identified in the IRAS sources, and thus contribute 
      significantly to the current star formation production. We speculate 
      that these clouds are the remnant fragments of a once large molecular 
      cloud that has since been dispersed.

\item The conjecture that the W4 \hii\ region as triggered star formation along
      the eastern edge of the W3 molecular cloud is re-investigated by 
      analyzing the spatial distribution of star forming regions. We find that 
      4 of the 5 massive star forming sites in the W3 GMC, as traced by 
      luminous IRAS point sources and/or radio continuum emission, are found 
      along the interface between the W4 \hii\ region and the W3 molecular 
      cloud. The western portion of the W3 cloud, despite having \about 75\% 
      of the molecular mass, contains only one known OB star forming region
      \citep{BKM00}. These characteristics are consistent with the long-held 
      notion that most of the massive star formation activity in the W3 
      molecular cloud has been induced by the expansion of the W4 \hii\ region 
      \citep{Lada78,Thronson80}.
\end{enumerate}

\acknowledgements

JMC acknowledges support from the James Clerk Maxwell Telescope Fellowship
while part of this work was carried out, and current support from the Owens 
Valley Radio Observatory and Long Term Space Astrophysics Grant NAG5-8217. 
The Five College Radio Astronomy Observatory is operated with support from NSF 
grant 94--20159. This publication makes use of data products from the Two 
Micron All Sky Survey, which is a joint project of the University of 
Massachusetts and the Infrared Processing and Analysis Center, funded by the 
National Aeronautics and Space Administration and the National Science 
Foundation.

\clearpage
 
\begin{figure}
\caption
{
    The location of the 34 IRAS point sources meeting our selection criteria
    (see text) overlaid on a grey scale image of the $^{12}$CO(1-0) integrated
    intensity $(\int T_{\rm A}^* dv)$ between $-57$\kms\ and $-32$\kms\ from
    the FCRAO Outer Galaxy Survey (Heyer et al.~1998). Small circles represent
    IRAS point sources with far-infrared luminosities $L_{\rm FIR} <
    500\thinspace L_\odot$, medium sized represent circles sources with
    $500\thinspace L_\odot \le L_{\rm FIR} < 5000\thinspace L_\odot$, and
    large circles represent sources with $L_{\rm FIR} \ge 5000\thinspace
    L_\odot$. The dash rectangle indicates the area that was searched for
    IRAS point sources.
    (For the astro-ph submission, Figure 1 is submitted as a JPEG image. 
     See also http://astro.caltech.edu/$\sim$jmc/papers/w3)
    \label{w345_co}
}
\end{figure}

\clearpage
\begin{figure}
\caption
{
    Contour map of the $^{12}$CO(1-0) integrated intensity shown in
    Figure~\ref{w345_co} overlaid on the \lhi\ continuum image from the DRAO
    Galactic Plane Survey \citep{NTD97}. The contour level shown is for
    an integrated intensity of
    $\int T_{\rm A}^* dv$ = 12\thinspace K\thinspace km\thinspace s$^{-1}$.
    (For the astro-ph submission, Figure 2 is submitted as a JPEG image. 
     See also http://astro.caltech.edu/$\sim$jmc/papers/w3)
    \label{w345_21cm}
}
\end{figure}

\clearpage
\begin{figure}
\caption
{
    Molecular line, near-infrared, and star count images for the 32 IRAS
    sources mosaicked at $K'$ band. For each source, the left panel is an
    image of the $^{12}$CO(1-0) integrated intensity $(\int T_{\rm A}^* dv)$
    from the FCRAO Outer Galaxy survey (Heyer et al.~1988) integrated over a
    $\pm$ 5\kms\ interval around the velocity toward the IRAS source. Each
    \co\ image is shown over a 30$' \times$ 30$'$ area centered on the
    IRAS point source position with a logarithmic intensity scale from
    log$_{10}$(1.0\kkms) to log$_{10}$(60.0\kkms). The adjacent panels show an
    image of the $^{13}$CO(1-0) integrated intensity centered on the IRAS
    source, the $K'$ band mosaic, and a $K'$ band stellar surface density map
    for stars with magnitudes between 11.5$^{\rm m}$ and 17.5$^{\rm m}$. The
    contour levels for the \thco\ maps begin at 1.0\kkms\ ($\int {\rm T_{MB}}
    dv$) with intervals of 2.0\kkms, except for IRAS 02232+6138, 02245+6115,
    02310+6133, 02455+6034, 02459+6029, 02531+6032, 02570+6028, 02593+6016,
    and 02575+6017, which have intervals of 8.0\kkms. No \thco\ data are
    present in any of the maps for $\Delta\delta \ge 1.5'$. The contours
    in the stellar surface density images begin at 2$\sigma$ above the mean
    derived stellar surface density (see text), with contours intervals of
    3$\sigma$. (For the astro-ph submission, Figure 3 is submitted as JPEG 
    images. See also http://astro.caltech.edu/$\sim$jmc/papers/w3)
    \label{images}
}
\end{figure}

\clearpage

\clearpage

\clearpage

\begin{deluxetable}{ccr@{\extracolsep{15pt}}cccl}
\scriptsize
\tablewidth{0pt}
\tablecaption{IRAS Sources in the W3/W4/W5 Region\label{tbl:iras}}
\tablehead{
\colhead{IRAS}                & 
\multicolumn{2}{c}{Galactic}  & 
\multicolumn{2}{c}{Equatorial (J2000)}  & 
\colhead{$V_{\rm LSR}$} & 
\colhead{Identification}\\
\cline{2-3}
\cline{4-5}\\
\colhead{}          &
\colhead{$\ell$}    &
\multicolumn{1}{c}{$b$} &
\colhead{$\alpha$}  &
\colhead{$\delta$}  &
\colhead{[km s$^{-1}$]}
}
\startdata
01546+6319\tablenotemark{a} & 130.2939 &  1.6549 &  01:58:19.5 & +63:33:59.1 & -54 & Weinberger 17\\
02008+6324\tablenotemark{a} & 130.9428 &  1.9137 &  02:04:34.3 & +63:38:31.5 & -45\\
02044+6031\tablenotemark{a} & 132.1562 & -0.7246 &  02:08:04.7 & +60:46:01.5 & -56 & AFGL 5066\\
02048+5957\tablenotemark{a} & 132.3671 & -1.2572 &  02:08:27.0 & +60:11:45.6 & -33\\
02054+6011\tablenotemark{a} & 132.3684 & -1.0212 &  02:09:01.3 & +60:25:16.3 & -57\\
02081+6225\tablenotemark{b} & 132.0137 &  1.2156 &  02:11:49.7 & +62:39:39.9 & -54\\
02175+5845                  & 134.2729 & -1.8974 &  02:21:07.7 & +58:59:06.3 & -48 & Weinberger 18\\
02186+6053\tablenotemark{a} & 133.6888 &  0.1617 &  02:22:22.2 & +61:07:11.3 & -50\\
02204+6128\tablenotemark{b} & 133.6961 &  0.7910 &  02:24:15.2 & +61:42:26.6 & -44\\
02220+6107\tablenotemark{a} & 134.3365 &  0.8897 &  02:29:35.0 & +61:34:07.8 & -51\\
02230+6202\tablenotemark{a} & 134.1165 &  1.7897 &  02:30:41.7 & +62:29:10.0 & -43 & G133.8+1.4; W3 N\\
02232+6138\tablenotemark{a} & 133.9434 &  1.0595 &  02:27:01.0 & +61:52:13.5 & -46 & W3(OH)\\
02245+6115\tablenotemark{a} & 134.2353 &  0.7485 &  02:28:21.5 & +61:28:29.0 & -49 & AFGL 333\\
02310+6133\tablenotemark{a} & 134.8298 &  1.3120 &  02:34:46.8 & +61:46:22.1 & -40\\
02327+6019\tablenotemark{a} & 135.5109 &  0.2576 &  02:36:33.5 & +60:32:10.2 & -43\\
02379+5724\tablenotemark{a} & 137.2764 & -2.1438 &  02:41:36.2 & +57:37:37.0 & -34\\
02407+6047\tablenotemark{a} & 136.2220 &  1.0813 &  02:44:37.7 & +60:59:52.7 & -44\\
02434+6018\tablenotemark{a} & 136.7205 &  0.7793 &  02:47:15.9 & +60:30:44.2 & -43\\
02439+6025\tablenotemark{a} & 136.7309 &  0.9202 &  02:47:50.2 & +60:38:05.5 & -37\\
02445+6042\tablenotemark{a} & 136.6723 &  1.2056 &  02:48:25.2 & +60:55:02.9 & -35\\
02455+6034                  & 136.8367 &  1.1367 &  02:49:23.2 & +60:47:01.0 & -40\\
02455+5808                  & 137.8969 & -1.0490 &  02:49:21.1 & +58:21:16.0 & -46\\
02459+6029                  & 136.9172 &  1.0850 &  02:49:47.6 & +60:42:06.9 & -40\\
02461+6147                  & 136.3845 &  2.2690 &  02:50:09.2 & +61:59:57.9 & -44 & AFGL 5085\\
02484+6022                  & 137.2467 &  1.1153 &  02:52:18.7 & +60:34:58.5 & -57\\
02495+6043                  & 137.2148 &  1.4911 &  02:53:27.8 & +60:55:58.1 & -37\\
02497+6217                  & 136.5388 &  2.8947 &  02:53:44.4 & +62:29:24.4 & -48\\
02511+6023                  & 137.5407 &  1.2788 &  02:55:03.0 & +60:35:44.3 & -38\\
02531+6032\tablenotemark{a} & 137.6937 &  1.5203 &  02:57:04.0 & +60:44:22.3 & -40\\
02541+6208\tablenotemark{a} & 137.0682 &  3.0016 &  02:58:13.2 & +62:20:29.0 & -51\\
02570+6028\tablenotemark{a} & 138.1514 &  1.6881 &  03:01:00.7 & +60:40:20.3 & -39\\
02572+6006\tablenotemark{a} & 138.3477 &  1.3738 &  03:01:11.7 & +60:18:08.7 & -38\\
02575+6017\tablenotemark{a} & 138.2913 &  1.5528 &  03:01:29.2 & +60:29:11.8 & -39 & AFGL 4029\\
02593+6016\tablenotemark{a} & 138.4977 &  1.6409 &  03:03:17.9 & +60:27:52.2 & -39 & Sh 201
\enddata
\vspace{-0.25truein}
\tablenotetext{a}{Coordinates of $K'$ mosaic registered using the 2MASS Image Atlas}
\tablenotetext{b}{Source not imaged at $K'$ band}
\end{deluxetable}

\clearpage
\begin{deluxetable}{crcccrc@{ $\pm$\extracolsep{1pt}}rr@{\extracolsep{5pt}}c}
\scriptsize
\tablewidth{335pt}
\tablecaption{Clusters Properties\label{tbl:clusters}}
\tablehead{
\colhead{IRAS}          &
\colhead{L$_{\rm FIR}$} &
\colhead{SpT}           &
\colhead{Ref.\tablenotemark{e}} &
\colhead{R$_{\rm eff}$} &
\colhead{N$_{\rm stars}$} &
\multicolumn{2}{c}{N$_{\rm cluster}$}\\
\colhead{}            &
\colhead{(L$_\odot$)} &
\colhead{}            &
\colhead{}            &
\colhead{(pc)}
}
\startdata
01546+6319                  &   260 & B1 & 1 & 0.54 &   89 &   54 &   6\\
02008+6324                  &     9 & \nodata & \nodata & \nodata & \multicolumn{2}{c}{\nodata}\\
02044+6031                  &  2900 & B0 & 1 & 0.73 &  204 &  147 &   8\\
02048+5957                  &   120 & \nodata & \nodata & 0.56 &   94 &   58 &   6\\
02054+6011                  &    91 & \nodata & \nodata & 0.59 &  104 &   70 &   6\\
02175+5845                  &   220 & \nodata & \nodata & 0.73 &  168 &  109 &   8\\
02186+6053                  &   120 & \nodata & \nodata & \nodata & \nodata & \multicolumn{2}{c}{\nodata}\\
02220+6107                  &    98 & \nodata & \nodata & \nodata & \nodata & \multicolumn{2}{c}{\nodata}\\
02230+6202                  & 16000 & O7 & 1 & \nodata & \nodata & \multicolumn{2}{c}{\nodata}\\
02232+6138\tablenotemark{a} & 46000 & B0.5 & 1 & 0.91 &  281 &  205 &   9\\
02245+6115                  &  3300 & B0.5 & 1 & 0.64 &  151 &  121 &   6\\
02310+6133                  &   270 & \nodata & \nodata & \nodata & \nodata & \multicolumn{2}{c}{\nodata}\\
02327+6019                  &   270 & \nodata & \nodata & 0.32 &   30 &   20 &   3\\
02379+5724                  &    50 & \nodata & \nodata & \nodata & \nodata & \multicolumn{2}{c}{\nodata}\\
02407+6047                  &   140 & \nodata & \nodata & 0.46 &   72 &   50 &   5\\
02434+6018                  &   770 & \nodata & \nodata & 0.37 &   39 &   24 &   4\\
02439+6025                  &    52 & \nodata & \nodata & \nodata & \nodata & \multicolumn{2}{c}{\nodata}\\
02445+6042                  &   420 & \nodata & \nodata & 0.45 &   41 &   23 &   4\\
02455+6034                  &  1100 & \nodata & \nodata & \nodata & \nodata & \multicolumn{2}{c}{\nodata}\\
02455+5808                  &    36 & \nodata & \nodata & \nodata & \nodata & \multicolumn{2}{c}{\nodata}\\
02459+6029                  &  2000 & \nodata & \nodata & 0.45 &   48 &   31 &   4\\
02461+6147                  &  2300 & B2 & 3 & 0.72 &  178 &  115 &   8\\
02484+6022                  &    88 & \nodata & \nodata & 0.62 &  140 &   86 &   7\\
02495+6043                  &    77 & \nodata & \nodata & \nodata & \nodata & \multicolumn{2}{c}{\nodata}\\
02497+6217                  &   140 & \nodata & \nodata & 0.38 &   50 &   36 &   4\\
02511+6023                  &   210 & B0.5 & 1 & \nodata & \nodata & \multicolumn{2}{c}{\nodata}\\
02531+6032                  &  1200 & B0.5 & 1 & \nodata & \nodata & \multicolumn{2}{c}{\nodata}\\
02541+6208                  &   360 & \nodata & \nodata & 0.45 &   62 &   40 &   5\\
02570+6028                  &   530 & \nodata & \nodata & 0.62 &  114 &   78 &   6\\
02572+6006                  &    72 & \nodata & \nodata & \nodata & \nodata & \multicolumn{2}{c}{\nodata}\\
02575+6017\tablenotemark{bcd} &  5700 & B1 & 2,4 & 1.00 &  340 &  240 &  10\\
02593+6016\tablenotemark{d} &  5800 & O9.5 & 1 & 0.62 &  127 &   88 &   6
\enddata
\vspace{-0.2truein}
\tablenotetext{a}{Observed at $K$ band by Tieftrunk et al. (1998)}
\tablenotetext{b}{Observed at $K$ band by Deharveng et al. (1997)}
\tablenotetext{c}{Observed at $K$ band by Hodapp (1994)}
\tablenotetext{d}{Observed at $K$ band by Carpenter et al. (1993)}
\tablenotetext{e}{References: 
    (1) Condon et al. 1998;
    (2) Kurtz, Churchwell, \& Wood 1994
    (3) McCutcheon et al. 1991
    (4) Carpenter, Snell, \& Schloerb 1991
}
\end{deluxetable}

\clearpage
\begin{deluxetable}{crr}
\tablewidth{0pt}
\tablecaption{Molecular Cloud Properties\label{tbl:clouds}}
\tablehead{
\colhead{IRAS}  & \multicolumn{1}{c}{Radius} & \multicolumn{1}{c}{Mass}\\
                & \multicolumn{1}{c}{(pc)}   & \multicolumn{1}{c}{(M$_\odot$)}
}
\startdata
01546+6319 &    0.49 &     27\\
02008+6324 &    0.60 &     64\\
02044+6031 &    0.49 &    140\\
02048+5957 &    0.32 &     29\\
02054+6011 &    0.51 &    100\\
02175+5845 &    0.68 &    410\\
02186+6053 & $>$  0.89 & $>$  600\\
02220+6107 &    0.37 &     69\\
02230+6202 & $>$  1.06 & $>$  670\\
02232+6138 & $>$  0.99 & $>$ 2700\\
02245+6115 & $>$  1.02 & $>$ 3200\\
02310+6133 & $>$  0.77 & $>$  680\\
02327+6019 &    0.69 &    410\\
02379+5724 &    0.51 &     71\\
02407+6047 &    0.70 &    290\\
02434+6018 &    0.42 &    120\\
02439+6025 &    0.28 &     29\\
02445+6042 & $>$  1.03 & $>$  920\\
02455+5808 & $>$  0.47 & $>$   45\\
02455+6034 & $>$  0.69 & $>$  670\\
02459+6029 & $>$  0.80 & $>$  800\\
02461+6147 &    0.34 &     83\\
02484+6022 &    0.63 &     43\\
02495+6043 &    0.47 &     69\\
02497+6217 &    0.49 &     55\\
02511+6023 &    0.45 &    130\\
02531+6032 & $>$  0.47 & $>$  290\\
02541+6208 &    0.42 &     56\\
02570+6028 &    0.53 &    380\\
02572+6006 &    0.40 &     28\\
02575+6017 &    0.69 &    870\\
02593+6016 & $>$  0.51 & $>$  300
\enddata
\end{deluxetable}

\end{document}